\documentclass[12pt]{article}
\pdfoutput=1
\usepackage{color}
\usepackage{amssymb,amsmath}
\usepackage{epsf}
\usepackage{epsfig}
\usepackage{afterpage}
\usepackage{longtable}
\usepackage{caption}
\usepackage{cite}
\usepackage{latexsym,mathrsfs}
\usepackage{graphics}
\usepackage{url}
\usepackage{paralist}
\usepackage{booktabs}
\usepackage{pgfplots}
\pgfplotsset{compat=1.11}
\usepackage{tikz-feynman}
\usepackage{tikz}
\usetikzlibrary{arrows,shapes}
\tikzstyle{every picture}+=[remember picture]

\newcommand{\gev}{\, {\rm GeV}}

\newcommand{\bsi}{B_6^{(1/2)}}
\newcommand{\bei}{B_8^{(3/2)}}

\newcommand{\muLow}{{\mu}}

\def\epe{\varepsilon'/\varepsilon}
\newcommand{\beq}{\begin{equation}}
\newcommand{\eeq}{\end{equation}}
\newcommand{\be}{\begin{equation}}
\newcommand{\ee}{\end{equation}}
\newcommand{\bi}{\begin{itemize}}
\newcommand{\ei}{\end{itemize}}
\newcommand{\ba}{\begin{array}}
\newcommand{\ea}{\end{array}}
\newcommand{\beqa}{\begin{eqnarray}}
\newcommand{\eeqa}{\end{eqnarray}}
\newcommand{\bea}{\begin{eqnarray}}
\newcommand{\eea}{\end{eqnarray}}
\newcommand{\beqn}{\begin{eqnarray}}
\newcommand{\eeqn}{\end{eqnarray}}

\newcommand{\epsK}{\varepsilon_K}

\definecolor{red}{cmyk}{0,1,1,0.4}

\usepackage{fancyhdr}
\advance \headheight by 3.0truept       

\begin{document}

\begin{flushright}
\end{flushright}

\medskip

\begin{center}
{\LARGE\bf
\boldmath{$\epe$ in and beyond the Standard Model}}
\\[0.8 cm]
{\bf Jason~Aebischer
 \\[0.5 cm]}
{\small
Excellence Cluster Universe, Boltzmannstr.~2, 85748~Garching, Germany }
\end{center}

\vskip0.41cm

\abstract{%
\noindent
Estimates of the CP violating observable $\epe$ have gained some attention in the past few years. Depending on the long-distance treatment used, they exhibit up to $2.9\sigma$ deviation from the experimentally measured value. Such a deviation motivates the investigation of New Physics (NP) effects in the process $K\to\pi\pi$. In my talk I will review the Standard Model (SM) prediction for $\epe$, with a special focus on the Dual QCD approach. On the NP side, I will discuss a recent computation of the hadronic matrix elements of NP operators. Furthermore a master formula for BSM effects in $\epe$ is presented. Finally, a treatment of $\epe$ using the SM effective theory (SMEFT) will be discussed together with possible correlations to other observables.
}

\thispagestyle{empty}
\newpage
\setcounter{page}{1}

\section{Introduction}

CP violation in the Standard Model (SM) has first been measured in the Kaon sector. The CP violating parameter measured in the famous Cronin-Fitch experiment \cite{Christenson:1964fg} is $\epsK$, which describes the mixing between CP and mass eigenstates of the neutral Kaon system. The parameter $\epsK$  measures the so-called CP violation through mixing. On the other hand, Kaons can also decay through direct CP violation. This CP violating decay is parametrized by the quantity $\varepsilon'$. The ratio of the two CP violating parameters $\epe$, where we suppress $K$ in $\epsK$, is also accessible experimentally, namely through a confrontation of the $K_L\to\pi^+\pi^-$ and $K_L\to\pi^0\pi^0$ decay widths. It has been measured by the NA48 \cite{Batley:2002gn} and KTeV \cite{AlaviHarati:2002ye,
Abouzaid:2010ny} collaborations and leads to an experimental world average of

\hspace{0.1cm}

\begin{equation}\label{eq:SMexp}
  (\epe)_\text{exp}  = (16.6 \pm 2.3) \times 10^{-4}\,.
\end{equation}
\hspace{2cm}

\noindent
The SM estimates for this observable depend on the long-distance (LD) treatment used to compute the $K\to\pi\pi$ hadronic matrix elements. As can be seen from Tab.~\ref{tab:SMpred}, the SM prediction differs for the three types of LD approaches and consequently there is some controversy over which treatment to use. The results obtained with Lattice QCD (LQCD) inputs as well as the ones in the Dual QCD (DQCD) approach are in good agreement with each other and exhibit about a $2.9\sigma$ deviation from the experimental value in eq.~\eqref{eq:SMexp}. The Chiral Perturbation Theory ($\chi$PT) approach leads to a value consistent with the SM, however exhibiting large uncertainties. Moreover the lower part of the error is consistent with the values obtained using Lattice or DQCD and therefore the situation is not conclusive.

\noindent
Taking the discrepancy between the SM prediction and the experimental value for granted, it is interesting to study beyond the SM (BSM) effect that could explain such deviations. In the following section I will review the SM prediction for $\epe$ based on the DQCD approach. In Sec.~\ref{sec:BSMME} the computation of the BSM matrix elements relevant for $\epe$ is discussed. In Sec.~\ref{sec:Masterform} a master formula for BSM effects in $\epe$ is presented and in Sec.~\ref{sec:SMEFT} the relation between $\epe$ and the SM effective theory (SMEFT) is discussed, before I summarize in Sec.~\ref{sec:concl}.

\section{$\epe$ in the SM}
To describe $\epe$ in a model-independent way, we use the effective Hamiltonian of three quark flavours which generates a $\Delta S=1$ transition. It consists of local operators multiplied by their corresponding Wilson coefficients and can be written as follows \cite{Buras:1991jm,Buras:1993dy,Ciuchini:1992tj,Ciuchini:1993vr}:

\begin{align}\label{eq:SMham}
  \mathcal{H}_{\Delta S = 1}^{(3)} &
  = - \sum_i C_i(\muLow) \, O_i\,.
\end{align}
This Hamiltonian is invariant under the unbroken gauge-group $SU(3)_c\times U(1)_{\rm em}$ and contains all the fields lighter than the charm quark as dynamical degrees of freedom. The minus sign is chosen to be in accord with the SMEFT conventions.

In the SM, the sum in eq.~\eqref{eq:SMham} contains seven four-quark operators consisting of $(V\pm A)$ currents as well as the chromomagnetic operator. The four-quark operators are generated through tree-level and box diagrams containing a $W$ boson and a gluon, as well as from QCD and Electroweak (EW) penguin diagrams. The seven effective operators can be written as linear combinations of the following vector-vector operators:

\begin{table}[tbp]
  \centering
  \begin{tabular}{cccc}
  \toprule
    Long-distance & SM prediction &
    Group &
    Ref.
    \\
    \midrule
  Lattice & $(1.4 \pm 6.9) \times 10^{-4}$ &
    RBC-UKQCD & \cite{Blum:2015ywa,Bai:2015nea}\\
    & $(1.9 \pm 4.5) \times 10^{-4}$ &
    Buras/Gorbahn/Jamin/J\"ager & \cite{Buras:2015yba}\\
    & $(1.1 \pm 5.1) \times 10^{-4}$ &
    Kitahara/Nierste/Tremper & \cite{Kitahara:2016nld}\\
    \midrule
    DQCD & $<(6.0\pm 2.4) \times 10^{-4}$  &
    Buras/G\'erard & \cite{Buras:2015xba}\\
    &$\qquad$ if $B_6<B_8 =B_8 \,(\text{LQCD})$ & &\\
    \midrule
    $\chi$PT
    & $(15 \pm 7) \times 10^{-4}$ &
    Gisbert/Pich & \cite{Gisbert:2017vvj}\\
    \bottomrule
  \end{tabular}
  \captionsetup{width=0.9\linewidth}
  \caption{SM estimates for $\epe$, using different treatments of the long-distance effects.
  }\label{tab:SMpred}
\end{table}

\begin{align}\label{eq:vecops}
  O_{VAB}^q &
  = (\bar s^i \gamma_{\mu} P_A d^i) (\bar q^j \gamma^{\mu} P_B q^j) \,,
&
  \widetilde{O}_{VAB}^q &
  = (\bar s^i \gamma_{\mu} P_A d^j) (\bar q^j \gamma^{\mu} P_B q^i) \,,
\end{align}
\hspace{2cm}

\noindent
where $P_{A,B}$ $(A,B=L,R)$ denote the chirality projection operators, $i,j$ are colour indices and $q=u,d,s$. The chromomagnetic operator reads:

\begin{align}\label{eq:chromo}
  O_{8g}      &
  = m_s(\bar s \, \sigma^{\mu\nu} T^A P_{L} d) \, G^A_{\mu\nu} \,,
\end{align}
\hspace{2cm}

\noindent
with $\sigma^{\mu\nu}=\frac{i}{2}[\gamma^{\mu},\gamma^{\nu}]$, $\,T^A$ being the $SU(3)_c$ generators and $G^A_{\mu\nu}$ the gluonic field-strength tensor.

Having the Hamiltonian of eq.~\eqref{eq:SMham} at hand allows to compute the $\epe$ observable, which is given by:

\begin{align}\label{eq:epspr}
  \frac{\varepsilon'}{\varepsilon} &
  = -\frac{\omega}{\sqrt{2}|\epsK|}
    \left[ \frac{\text{Im}A_0}{\text{Re}A_0}
         - \frac{\text{Im}A_2}{\text{Re}A_2} \right]\,.
\end{align}
\newline

\noindent
Here $\omega = {\text{Re}A_2}/{\text{Re}A_0} \approx 1/22$, reflecting the $\Delta I =1/2$ rule, and $\epsK$ is the Kaon mixing parameter mentioned before. The expression is therefore determined by the isospin amplitudes $A_{0,2}$ defined by

\begin{align}
  A_{0,2} &
  = \Big\langle (\pi\pi)_{I=0,2}\, \Big|\; \mathcal{H}_{\Delta S = 1}^{(3)}(\muLow)
  \;\Big|\, K \Big\rangle \,.
\end{align}
\hspace{2cm}

\noindent
After having fixed the Wilson coefficients of $\mathcal{H}_{\Delta S = 1}^{(3)}$ by performing a matching procedure, the only remaining task is to compute the hadronic matrix elements of the local operators in eq.~\eqref{eq:SMham}. In the following subsection, we will look into this computation by employing the DQCD approach.

\subsection{Long-distance effects in the DQCD approach}
The DQCD is based on the large $N_c$ limit, first studied by t'Hooft \cite{'tHooft:1973jz,'tHooft:1974hx} and Witten \cite{Witten:1979kh,Treiman:1986ep} for strong interactions.
To study hadronic weak decays, the following truncated Chiral Lagrangian is used \cite{Bardeen:1986vp,Bardeen:1986uz,Bardeen:1986vz}:
\begin{equation}\label{eq:chiralLag}
\mathcal{L}_{tr}=\frac{F^2}{8}\left[\text{Tr}(D^\mu UD_\mu U^\dagger)+r\text{Tr}(mU^\dagger+\text{h.c.})-\frac{r}{\Lambda^2_\chi}\text{Tr}(mD^2U^\dagger+\text{h.c.})\right] {\,,}
\end{equation}
with the unitary chiral matrix and the octet of lowest-lying pseudoscalars
\begin{equation}
U=\exp(i\sqrt{2}\frac{\Pi}{F}), \qquad
\Pi=\sum_{\alpha=1}^8\lambda_\alpha\pi^\alpha{\,.}
\end{equation}
The Lagrangian depends on the quark mass matrix and the chiral enhancement factor

\begin{equation}
  m=\text{diag}(m_u,m_d,m_s)\,, \qquad r=\frac{2m_K^2}{m_s^2+m_d^2}\,.
\end{equation}
\hspace{2cm}

\noindent
It contains a hadronic mass scale $\Lambda_\chi$ corresponding to higher resonances.

\noindent
Employing now the large $N_c$ limit, the Lagrangian of eq.~\eqref{eq:chiralLag} can be matched onto the regular QCD Lagrangian containing quark and gluon fields only. In the chiral limit and at order $\mathcal{O}(p^2)$ the quark currents are then given by:

\begin{equation}\label{eq:mesrep}
(\gamma^\mu P_L)^{ba}=i \frac{F^2}{4} (\partial^\mu U U^\dagger)^{ab}, \qquad (P_L)^{ba}=- \frac{F^2}{8} r (U)^{ab}\,,  \qquad (\sigma^{\mu\nu}P_L)^{ab} = 0\,,
\end{equation}
\hspace{2cm}

\noindent
for the flavour indices $a,b$. The chirality flipped versions are obtained by the replacement $U\leftrightarrow U^\dag$. These relations allow to express the local operators in terms of the lowest-lying mesons and therefore to compute their corresponding matrix elements. Furthermore, this framework allows to study the renormalization group (RG) evolution of the matrix elements up to a scale of $\mathcal{O}(1\text{GeV})$ until where the theory is valid. This RG evolution is dubbed meson evolution.

\noindent
The DQCD approach was first employed in the context of $K\to\pi\pi$ matrix elements in \cite{Bardeen:1986vp,Bardeen:1986vz,Buras:2014maa}. Its validity is confirmed by results obtained within LQCD. Among them is the correctly predicted hierarchy of the bag factors for the SM operators $Q_6$ and $Q_8$ \cite{Buras:2015xba}

\begin{equation}\label{BG}
  \bsi \leq \bei < 1 \, .
\end{equation}
\hspace{2cm}

\noindent
Also the explicit calculations for $B_6^{(1/2)}(m_c),\,B_8^{(3/2)}(m_c)$ are in good agreement with the Lattice results \cite{Bai:2015nea,Blum:2015ywa}. Not only for the SM four-quark operators but also for the matrix element of the chromomagnetic operator of eq.~\eqref{eq:chromo}, DQCD \cite{Buras:2018evv} agrees well with LQCD \cite{Constantinou:2017sgv}. Furthermore, the impact of final state interactions has been analysed within the DQCD approach in \cite{Buras:2016fys} and has been shown to be less important for  $\epe$ than for the $\Delta I=1/2$ rule, and less important than meson evolution which is responsible for (\ref{BG}).
Finally DQCD also  allows, with the help of meson evolution, to understand
the pattern of the BSM $K^0-\bar K^0$ mixing matrix elements \cite{Buras:2018lgu} obtained by  LQCD \cite{Carrasco:2015pra,Jang:2015sla,Boyle:2017ssm}.
 More information on DQCD can be found in the original papers and in the reviews in \cite{Buras:2018hze,Buras:2014maa}.

\section{BSM matrix elements for $\epe$}\label{sec:BSMME}
Generalizing the SM Hamiltonian by allowing for all possible Lorentz- and gauge invariant operators, one finds that there are 13 additional four-quark operators to be added to $\mathcal{H}_{\Delta S = 1}^{(3)}$. Three of them are vector-vector operators which are independent of the seven operators generated within the SM. They can also be written as linear combinations of the operators in eq.~\eqref{eq:vecops}. The other BSM operators consist of scalar or tensor bilinears and can be written as linear combinations of the following operators:

\begin{align}
  O_{SAB}^q &
  = (\bar s^i P_A d^i) (\bar q^jP_B q^j) \,,
&
  \widetilde{O}_{SAB}^q &
  = (\bar s^i  P_A d^j) (\bar q^j  P_B q^i) \,, \\
  O_{TA}^q &
  = (\bar s^i \sigma_{\mu\nu} P_A d^i) (\bar q^j\sigma^{\mu\nu}P_A q^j) \,,
&
  \widetilde{O}_{TA}^q &
  = (\bar s^i \sigma_{\mu\nu} P_A d^j) (\bar q^j\sigma^{\mu\nu}P_A q^i) \,,
\end{align}
\hspace{2cm}

\noindent
for $q=u,d,s$. Two equivalent bases for the 13 BSM operators can be found in \cite{Aebischer:2018rrz}.

\noindent
The $K\to\pi\pi$ matrix elements of these BSM operators have been calculated for the first time in \cite{Aebischer:2018rrz}, using the DQCD approach. They were computed first at the factorization scale $\mu_F$ at which the meson representation of eq.~\eqref{eq:mesrep} holds. The factorization scale corresponds to very low momenta of $\mathcal{O}(p^2\approx 0)$. Since the observable $\epe$ is usually computed at the charm scale $\mu_c=\mathcal{O}(m_c)$, the running of the matrix elements has to be performed from the factorization scale up to the scale $\mu_c$ via the meson evolution for scales below $1\gev$ followed by the
usual QCD evolution.

The explicit expressions and numerical values of all the matrix elements at the charm scale as well as further details of the computation can be found in \cite{Aebischer:2018rrz}. Here, we summarize only quantitatively the results of the analysis. For the different types of BSM operators, one finds for their respective matrix elements at the factorization scale $\mu_F$ and at the charm scale $\mu_c$:
\newline

\begin{itemize}
\item Vector operators: small at $\mu_F$ and at $\mu_c$.
\item Scalar operators: large at $\mu_F$, moderate at $\mu_c$.
\item Tensor operators: zero at $\mu_F$, large at $\mu_c$.
\item Scalar/Tensor operators containing three $s$ quarks: zero at $\mu_F$ and at $\mu_c$.
\end{itemize}

\section{Master formula for BSM effects in $\epe$}\label{sec:Masterform}
Knowing the matrix elements for the complete set of local effective operators relevant for $\epe$ allows for a model-independent analysis of the BSM effects. In this section we provide the means for such an analysis in the form of a master formula for $\epe$ \cite{Aebischer:2018quc}. For this purpose, we split the observable in the following way:
\begin{align}
  \frac{\varepsilon'}{\varepsilon} &
  = \left(\frac{\varepsilon'}{\varepsilon}\right)_\text{SM}
  + \left(\frac{\varepsilon'}{\varepsilon}\right)_\text{BSM} \,,
\end{align}
and focus on the BSM part. Since many NP scenarios contain heavy degrees of freedom with a mass scale above the EW scale, it is reasonable to provide a master formula evaluated at the EW scale $\mu_W$. Consequently, a NP analysis of a particular model only requires a simple tree-level matching at $\mu_W$. To evaluate eq.~\eqref{eq:epspr} at the EW scale, the RG evolution of the matrix elements from $\mu_c$ up to $\mu_W$ has to be taken into account \cite{Aebischer:2017gaw,Jenkins:2017dyc}. In the running up to the EW scale new operators containing $c$ and $b$ quarks will be generated through QCD and QED mixing, leading to the more general Hamiltonian of five flavours $\mathcal{H}_{\Delta S = 1}^{(5)}$. The master formula will therefore depend on the Wilson coefficients of all such effective operators. Setting the parameter $\epsK$ as well as $\text{Re}(A_0)$ and $\text{Re}(A_2)$ appearing in eq.~\eqref{eq:epspr} to their experimental values \cite{Cirigliano:2011ny} one finds the following master formula:

\begin{align}
  \label{eq:master}
  \left(\frac{\varepsilon'}{\varepsilon}\right)_\text{BSM} &
  = \sum_i  P_i(\mu_W) ~\text{Im}\left[ C_i(\mu_W) - C^\prime_i(\mu_W)\right]
  \times (1\,\text{TeV})^2,
\end{align}

with

\begin{align}
  \label{eq:master2}
  P_i(\mu_W) & = \sum_{j} \sum_{I=0,2} p_{ij}^{(I)}(\mu_W, \mu_c)
  \,\left[\frac{\langle O_j (\mu_c) \rangle_I}{\text{GeV}^3}\right].
\end{align}

Here, the $p_{ij}^{(I)}$ contain the evolution from $\mu_c$ to $\mu_W$. The matrix elements $\langle O_j (\mu_c) \rangle_I$ are taken from LQCD \cite{Blum:2015ywa,Bai:2015nea} for the SM operators and from DQCD \cite{Aebischer:2018rrz} for the BSM operators. The crucial objects determining the impact of each Wilson coefficient on $\epe$ are the $P_i$ values. These were obtained using the public codes \texttt{wcxf} \cite{Aebischer:2017ugx} for the basis change, \texttt{wilson} \cite{Aebischer:2018bkb} for the RG running and \texttt{flavio} \cite{Straub:2018kue} to compute $\epe$ at the EW scale. The $P_i$ values of the full set of operators contained in $\mathcal{H}_{\Delta S = 1}^{(5)}$ can be grouped into five classes $(A-E)$, which are listed in Tab.~\ref{tab:Pis}. The operators either give a direct BSM contribution to $\epe$ through their matrix element (ME) or contribute to the observable indirectly through RG mixing. For further details and the explicit values of the $P_i$'s as well as their respective uncertainties we refer to \cite{Aebischer:2018quc}.

\begin{table}
\centering
\renewcommand{\arraystretch}{1.4}
\begin{tabular}{ccccc}
\hline
  Class &  Type & $O_i$       & $P_i$ & Impact  \\
  \hline

  A & SM
& $O_{VAB}^{u,d}, \widetilde O_{VAB}^{u,d}, O_{SLR}^{d}$                                       & can be large &      ME \\
 &
& $O_{VAB}^{s,c,b}, \widetilde O_{VAB}^{c,b}, O_{SLR}^{s}$                                       & small &      Mixing \\

\hline
  B & Chromomagnetic      & $O_{8g}$                & small  &    Mixing     \\
& Scalar: $s ,c, b$       & $O_{SLL}^{s,c,b}, \widetilde O_{SLL}^{c,b} $        & small  &    Mixing     \\

& Tensor: $s ,c, b$       & $O_{TLL}^{s,c,b}$        & small  &    Mixing     \\
\hline
C & Scalar: $u$       & $O_{SLL}^{u}, \widetilde O_{SLL}^{u} $        & small  &    ME     \\
& Tensor: $u$       & $O_{TLL}^{u}, \widetilde O_{TLL}^{u} $        & large  &    ME     \\
\hline
D & Scalar: $d$       & $O_{SLL}^{d} $        & small  &    ME     \\
& Tensor: $d$       & $O_{TLL}^{d} $        & large  &    ME     \\
\hline
E & Scalar LR: $u$       & $O_{SLR}^{u}, \widetilde O_{SLR}^{u} $        & can be large  &    ME     \\
\hline
\end{tabular}
\captionsetup{width=0.9\linewidth}
\caption{$P_i$ values of the effective operators relevant for $\epe$ at the EW scale, grouped into five classes (A-E). The operators either contribute via their matrix element (ME) or through mixing effects to the observable.}
\label{tab:Pis}
\end{table}

\section{$\epe$ meets SMEFT}\label{sec:SMEFT}
Assuming that NP manifests itself at scales much higher than the EW scale, the SMEFT \cite{Buchmuller:1985jz,Grzadkowski:2010es} consists of a valid low-energy effective theory of such a NP scenario. Therefore it is reasonable to adopt the SMEFT as an intermediate theory between any NP model and the SM. This procedure allows to describe NP effects in a model independent way. The complete tree-level matching of the SMEFT onto the weak effective theory is done in \cite{Aebischer:2015fzz,Jenkins:2017jig} and in \cite{Aebischer:2018csl} all the SMEFT operators relevant for $\epe$ have been identified. There are:

\begin{itemize}
  \item vector four-quark operators: $\mathcal{O}_{qq}^{(1,3)}, \mathcal{O}_{qu}^{(1,8)},\mathcal{O}_{qd}^{(1,8)}, \mathcal{O}_{ud}^{(1,8)} ,\mathcal{O}_{dd}\,,$
  \item scalar four-quark operators: $\mathcal{O}_{quqd}^{(1,8)}\,,$
  \item modified W and Z couplings: $\mathcal{O}_{Hq}^{(1,3)},\mathcal{O}_{Hd},\mathcal{O}_{Hud}\,,$
  \item chromomagnetic dipole operator: $\mathcal{O}_{dG}$.
\end{itemize}
An effect in $\epe$ stemming from SMEFT operators can result in correlations with other observables. This occurs for operators containing a quark doublet after changing from the flavour to the interaction basis, or through flavour dependent RG mixing effects. In \cite{Aebischer:2018csl} correlations of $\epe$ to $\Delta S =2$ and $\Delta C=1, 2$ processes, semileptonic Kaon decays, the electroweak $T$ parameter, collider constraints as well as the neutron electric dipole moment (EDM) have been analysed. Furthermore, several tree-level mediator scenarios have been studied, which are summarised in Tab.~\ref{tab:treemed}. Further details on correlations of $\epe$ and the observables mentioned here can be found in \cite{Aebischer:2018csl}.

\section{Summary}\label{sec:concl}

The hadronic matrix elements for the BSM operators relevant for $\epe$ have been presented for the first time in \cite{Aebischer:2018rrz}. The newly acquired matrix elements allowed for the first time to derive a master formula for $\epe$, depending on SM and BSM operators. This master formula is presented in \cite{Aebischer:2018quc} and is already included in several public codes, such as \texttt{flavio} \cite{Straub:2018kue} and \texttt{smelli} \cite{Aebischer:2018iyb}. Based on this master formula, different correlations of $\epe$ to other observables have been analysed in the context of the SMEFT in \cite{Aebischer:2018csl}.

\begin{table}[tbp]
\centering
\renewcommand{\arraystretch}{1.4}
\begin{tabular}{cccc}
\hline
  Mediator &  SM Representation & SMEFT       & Correlation   \\
  \hline

  $Z'$ & $(1,1)_0$ & $ \mathcal{O}_{qd}^{(1)}$  & $\epsK$   \\
   &  & $\mathcal{O}_{qu}^{(1)}$  &  $pp\rightarrow jj$  \\
      &  & $\mathcal{O}_{HD}$  &  T parameter  \\
\hline
  Coloured scalar & $(8,2)_{1/2}$ & $ \mathcal{O}_{qd}^{(1)}$  & $\epsK$   \\
     &  & $\mathcal{O}^{(8)}_{quqd}$  &  neutron EDM  \\
\hline
\end{tabular}
\captionsetup{width=0.9\linewidth}
\caption{Tree-level models, which can have a sizable effect in $\epe$ and their correlations to other observables.}
\label{tab:treemed}
\end{table}
\section*{Acknowledgements}
 It is a pleasure to thank the organizers of this workshop for inviting me
to this interesting event. In particular I would like to thank my great collaborators: Christoph Bobeth, Andrzej Buras, Jean-Marc
 G{\'e}rard and David Straub for a very nice and inspiring collaboration. A special thanks goes to Andrzej Buras for all his support and for giving me the opportunity to give this talk. This research was
supported by the DFG cluster of excellence ``Origin and Structure of the Universe''.

\bibliographystyle{JHEP}
\bibliography{Bookallrefs}
\end{document}